\documentclass[aps,prd,twocolumn,showpacs,10pt,superscriptaddress,preprintnumbers,nofootinbib]{revtex4-1}
\usepackage{graphicx}
\usepackage{dcolumn}
\usepackage{bm}
\usepackage{fancyhdr}
\usepackage{amsmath,amsfonts,extarrows,amssymb,mathrsfs,times,bm}
\usepackage{extarrows}
\usepackage{graphicx}
\usepackage{float}
\usepackage{graphicx,epstopdf}
\usepackage{dcolumn}
\usepackage{exscale}
\usepackage{relsize}
\usepackage{mathtools}
\usepackage{mhchem}
\usepackage[colorlinks=true,breaklinks=true,linkcolor=blue,citecolor=blue,urlcolor=blue]{hyperref}

\pdfoutput=1

\usepackage[export]{adjustbox}
\usepackage{xcolor}
\usepackage{graphicx}
\usepackage{dcolumn}
\usepackage{bm}

\begin{document}
\title{Scalar-Connection Gravity and Spontaneous Scalarization}
\author{Hemza Azri}
\email{hmazri@uaeu.ac.ae}
\affiliation{Department of Physics, United Arab Emirates University,
Al Ain 15551 Abu Dhabi, UAE}%
\author{Salah Nasri$^{1}$}
 \email{snasri@uaeu.ac.ae}
\affiliation{International Center for Theoretical Physics, Strada Costiera 11, I-34151 Trieste, Italy}%

\date{\today}

\begin{abstract}
Scalar-tensor theories of gravity are known to allow significant deviations from general relativity through various astrophysical phenomena. In this paper, we formulate a scalar-connection gravity by setting up scalars and connection configurations instead of metric. Since the matter sector is not straightforward to conceive without a metric, we invoke cosmological fluids in terms of their one-form velocity in the volume element of the invariant action. This leads to gravitational equations with a perfect fluid source and a generated metric, which are expected to produce reasonable deviations from general relativity in the strong field regime. As a relevant application, we study a spontaneous scalarization mechanism and show that the Damour-Esposito-Far\`{e}se model arises in a certain class of scalar-connection gravity. Furthermore, we investigate a general study in which the present framework becomes distinguishable from the famed scalar-tensor theories.
\end{abstract}

\maketitle

\section{Introductory remarks and motivations}
The significant advances on observational astrophysics in the last few decades have finally led to the detection of the gravitational waves, a promising prediction of general relativity (GR), and opened a new window towards testing alternative theories of gravity \cite{ligo}. Among the prominent theories is the so-called scalar-tensor gravity (STG) which has gained much attentions in the last few decades \cite{book}.  This type of theories is expected to reveal important and testable deviations from GR in  strong regimes by studying the properties of high-density objects such as neutron stars \cite{will}. Recently, a more general form of STG, namely the Horndeski  theory \cite{horndeski}, have gained much attention since it stands viable in astrophysics and cosmology beyond GR (see also \cite{posedness} for the problem of well-posedness of the same type of theories). On the other hand, scalar tensor theory in the context of Palatini, metric-affine, and teleparallel gravity has also been investigated \cite{palatini stg}. 

Although it is considered as an alternative theory of gravity, STG is  however  formulated through the strong foundations of GR as a geometric theory of spacetime. Given that the gravitational potential is represented by the components of the metric tensor $g_{\mu \nu}$ as a fundamental field, one may refer to GR then as a tensor theory.  Additionally, in STG  one  incorporates a dynamical scalar field $\phi$ so that    gravity is mediated by the scalar-tensor configuration $(\phi,g)$. Like any alternative or modified theory of gravity, STG also has some limit in which the physical predictions are consistent with those of GR, whereas for weak fields, it would correspond to the Newtonian regime. For instance, Brans-Dicke theory, a particular and a former type of STG, reduces to GR when the scalar field gains a constant value \cite{book, will}. However, in the general case, specific restrictions must be imposed for certain values of the coupling between the scalar field and the spacetime curvature so that the theory possesses a GR solution (see Ref \cite{GR limit} for more details.) The presence of one or multiple scalar field in this type of theories might be motivated by high-energy physics models that aim in unifying the fundamental interactions. 

On  the physics side, one of the most notable consequences arising in some specific STG's is spontaneous scalarization. Motivated by spontaneous magnetization phenomenon, certain class of STG has been proposed so that the solutions to the field equations of GR arising from vanishing scalar field  is unstable, i.e, a tiny perturbations around this solution can be amplified by high-density matter such as a neutron star and drive it to a new stable configurations. The STG action which is  the essence of this mechanism is written (in units $G_{N}=1$) as \cite{damour}
\begin{eqnarray}
\label{stg action}
S[\phi,g]= &&\frac{1}{16\pi} \int d^{4}x\sqrt{|g|}\, \Big[g^{\mu\nu}R_{\mu\nu}(g)-2g^{\mu\nu}\partial_{\mu}\phi\partial_{\nu}\phi \Big] \nonumber \\
&&+S_{m}\left[\psi_{m}, \Omega^{2}(\phi)g_{\mu\nu} \right],
\end{eqnarray}
where the first term is the Einstein-Hilbert action and the last term represents the action of matter fields $\psi_{m}$.

Although the scalar field is chosen to be massless (see \cite{fethi} for the massive case), the scaled metric appearing in the matter sector induces an effective potential within matter as\footnote{In the original paper \cite{damour}, instead of $\Omega^{2}(\phi)$, the notation $A^{2}(\phi)$ is used to rescale the metric. Here we use the former so that there will not be confusions with other parameters in the next sections.}
\begin{eqnarray}
\label{eom in STG}
\Box \phi =-4\pi \left(\frac{\partial (\ln \Omega)}{\partial \phi} \right)T,
\end{eqnarray}
with $T$ being the trace of the stress-energy tensor of matter. In \cite{damour}, the coupling function $\Omega(\phi)$ has been chosen as $\Omega(\phi)=e^{\beta \phi^{2}/2}$ where $\beta$ is a constant, so that the STG accepts GR solution at the vacuum $\phi=0$. It turns out that for negative values of the parameter $\beta$ this solution is unstable and a suitable conditions within a neutron star drive the system toward a new and stable configuration with nonzero scalar field. This phenomenon has received much attentions recently where it has been considered as one of the most important phenomenon that can lead to testable predictions on the physics of neutron stars beyond GR \cite{kunz}.     

Returning to the theory of gravity at hand, one notices that STG described by (\ref{stg action}) is based on the following requirements:
\begin{itemize}
    \item The gravitational sector of the theory is given in terms of the metric tensor from which arises  other geometric quantities, such as the volume measure and the curvature.
    \item The matter sector, on the other hand, is not coupled directly to this metric but to its scaled version. This leads to 
    the appearance of the well known Einstein and Jordan frames in the theory. 
\end{itemize}
However, generally speaking, the main geometric factor which is at the heart of relativistic gravity is the spacetime curvature. It is known that this quantity is constructed from a connection rather than the metric tensor, which  
can be reduced to a metric connection only \textit{a posteriori} at certain conditions. Thus, it is rather natural to consider a gravitational action alternative to that of STG, and which stands on the following properties:
\begin{enumerate}
    \item The gravitational sector is given in terms of an affine connection $\Gamma (x)$ from which arise all other geometric quantities.
    \item The matter sector is metricless, and hence there is no notion of Einstein-Jordan frame in the first place.
    \item As in STG, a dynamical scalar field contributes in mediating gravity.
\end{enumerate}
With these properties, the resulting gravitational action based on the configuration $(\phi,\Gamma)$ will describe a Scalar-Connection Gravity (SCG) which we will present and explore in details in this paper. We should emphasize here that the name "Scalar-Connection Gravity" is proposed in line with the main and new properties of this type of actions stated above (compared to its counterpart scalar-tensor-theories.)

Forming an action principle without introducing a metric \textit{a priori} cannot be trivial as in the philosophy of GR especially when matter is involved. This can be simply understood from the fact that, in the standard Lagrangian formalism, matter fields are placed through their covariant kinetic terms formed by the metric itself. However, several attempts have been made to overcome this difficulty and couple various sources of matter such as scalar and vector fields through the Ricci curvature \cite{kijowski}, and fermions by integrating out the metric \cite{metric free} (see also Ref \cite{affine dark matter} for scalar dark matter in this context).  For the case of scalar matter, a less challenging fields in a purely affine spacetime, it has been shown that the proposed Lagrangian densities though nonpolynomial produce the same field equations of GR with a propagating scalar fields. These Lagrangians are obtained by applying certain transformations analogous to Legendre transformation in classical mechanics by noticing that the metric appears as the canonical momentum conjugate to the connection in the Palatini formalism. Nevertheless, although this equivalence between GR and purely affine formulation is easily demonstrated when the scalars are minimally coupled to gravity, the presence of explicit interaction between curvature and matter produces considerable differences between the two formulations \cite{affine inflation}. For instance, it is known that due to the nonlinearity (in metric) of the gravitational action of GR, the presence of any nonminimal interaction induces anisotropic stresses which in turn source nonadiabatic perturbations. On the other hand, since the curvature is linear in the connection, these anisotropic sources do not arise in a purely affine formulation when matter is nonminimally coupled to gravity \cite{entropy production}. Concerning vector fields ($U(1)$ in particular), the proposed Lagrangian densities are such that the field strength is coupled to the affine curvature the same way it couples to metric in GR. In this case, the resulting nonlinear terms in the affine curvature lead to a complicated dynamical equation which impedes the smooth emergence of the metric as in the scalar models. Nonetheless, it has been shown that these actions can produce Einstein's field equations with a propagating Maxwell field with the aid of a local reference frame \cite{kijowski}. Last but not least, coupling fermions to gravity is more challenging than the other kind of matter. In fact, while the fermion field is a scalar under general transformation, the gamma matrices are vectors, and a flat metric is necessary for the spin transformation to make sense locally. That being said, the fact remains on how to include all types of matter  fields (vectors, scalar, and fermions) in a standard model where gravity is described only by an affine connection.  

In the present work, in addition to scalar fields, we intend to incorporate matter as a cosmological fluid by extending the simplest purely affine (Eddington's) theories. Despite some similarities in the structures, the essential difference between SCG and usual affine theories including scalar fields is the inclusion of matter as a perfect fluid. Additionally, in SCG, the scalar field will also mediate gravity.

With this new structure we will show that for a massless scalar, an induced effective potential can arise within matter not in terms of the trace of the energy-momentum tensor as in (\ref{stg action}) but through distinct proportions of both energy density and pressure of the perfect fluid. Hence, small perturbations of the scalar field around the unstable GR solution ($\phi=0$) is augmented by the effects of these induced quantities and the system then evolves to a stable configuration, after high  enough  growth, due to nonlinear interactions between matter and the scalar field. We will show that in a particular class of SCG, one deduces also the Damour-Esposito-Far\`{e}se model of spontaneous growth where the effective potential comes out within the trace of the stress-energy tensor of matter.  

The paper is organized as follows, in section \ref{sec: SCG} we discuss the motivation behind investigating a gravity based on connection instead of metric, and then construct the invariant action that will be at the heart of the proposed SCG. We then proceed to deriving the main field equations; the gravitational and scalar field equations of motion where we find that matter results through the energy-stress tensor of a cosmological fluid. In section \ref{sec:spontaneous scalarization} we apply this to spontaneous scalarization mechanism and conclude in section \ref{sec:concluding remarks}.

\section{Scalar-Connection gravity}
\label{sec: SCG}
In this section we present a scalar-connection theory of gravity based on the properties stated above. To that end, the first attempt is to set up the main formalism in which we incorporate matter as a cosmological fluid in a spacetime endowed with an affine connection as the sole fundamental geometric quantity. Although the metric is not necessary for describing spacetime curvature, thus gravity (see the discussions above), we know however that concepts of lengths and angles are meaningful in the observable universe. In SCG, the metric is an auxiliary quantity of secondary importance, but it arises from the theory itself.

With  the absence of metric tensor, the number of quantities that one could consider are less than in the metric case, recalling  that scalars formed by contractions (using metric) are not allowed in the first place. In the geometric sector, one could however consider vector velocities in addition to a symmetric connection $\Gamma_{\mu\nu}^{\lambda}(x)$ and the associated curvature or Ricci tensor $\textbf{R}_{\mu\nu}(\Gamma)$.  Although the affine connection and its curvature can be, in general, taken asymmetric, in the present formulation we will be dealing with only the symmetric cases that are sufficient for a geometric description of the gravitational phenomenon. 
Nonsymmetric affinity on the other hand implies the appearance of an additional field in any affine theory of gravity, namely a torsion tensor, that can be appropriate for describing a nongravitational phenomena. In fact, it is shown that theories with asymmetric connections can be treated as an affine theory based on symmetric connection interacting with an additional matter fields \cite{nonsymmetric} However, it should be noticed here that considering a symmetric connection from the beginning, as we do here, does not require that it is a metric (Levi-Civita) connection in the first place, and the compatibility relation (see below) will not be imposed from scratch but an outcome of the theory.

Despite its simple structure, forming the  familiar polynomial gravitational action in this pure affine background is a perplexing problem, and several attempts have been made in this respect \cite{oscar}. Nevertheless, the SCG invariant action can be  formed in terms of a volume measure as follows
\begin{widetext}
\begin{equation}
\label{min_action}
S\left[\phi, \Gamma \right]=
\int d^{4}x \frac{\sqrt{| \kappa^{-1}\left[\textbf{R}_{\mu\nu}(\Gamma) -\nabla_{\mu}\phi\nabla_{\nu}\phi\right] -\mathcal{A}(x,\phi)\textbf{u}_{\mu}\textbf{u}_{\nu} |}}{\mathcal{\mathcal{B}}(x,\phi)}
\end{equation}
\end{widetext}
where the symbol $|.|$ refers to the absolute value of determinant and $\kappa=8\pi G_{\text{N}}$ is the gravitational constant. The parameters that appear in the action are as follows:  
\begin{itemize}
    \item $\textbf{u}_{\mu}$ is a one-form that will be associated to vector velocities of the fluids.
    \item $\mathcal{A}(x,\phi)$ refers to kinetic part of the fluid. A sufficient property for the emergence of the stress-energy tensor of the perfect fluid from this action (see the derivation below), is that this quantity is taken a scalar and independent of other geometric entities. Coupling this term to the connection can be understood implicitly since both curvature (that involves the connection) and the kinetic term of matter contribute to the total expression of the determinant. In case a connection-dependence is further imposed on $\mathcal{A}(x,\phi)$, one would deal then with the existence of an additional dynamical attribute of matter, namely the hypermomentum tensor \cite{hypermomentum}. Studying the latter case puts it beyond the scope of the present work.   
    \item $\mathcal{B}(x,\phi)$ is a scalar and it refers to "potential" part of the fluids. An essential requirement here is that $\mathcal{B}(x,\phi)\neq 0$ for a well defined action. As we will see later on, this stands important in generating the metric tensor through the dynamical equation obtained from this action.
\end{itemize}
Additionally, the action contains a dynamical scalar field $\phi$. Here we are refereeing to  the fluid parameters $\mathcal{A}(x,\phi)$ and $\mathcal{B}(x,\phi)$ as kinetic and potential only  by analogy with the scalar field. The above  action 
enjoys the general covariance, and it is expected to lead to perfect fluids in terms of energy density and pressure at cosmological scales. In general, deriving fluid dynamics from principle of variation has been interesting in their own where one can formulate an Eulerian relativistic hydrodynamics \cite{schutz}. In this work we will not be interested in an Eulerian analogues of fluids in affine spacetime. Nevertheless, since the matter sector is
not trivially conceived without metric in an action principle, we intend here to set up an indirect picture of matter which will manifest as a cosmological fluid and leads to possibly interesting physics as we shall see in the subsequent sections on spontaneous scalarization.  

Now, requiring the action to be stationary under variation   with respect to the symmetric connection yields  
    \begin{eqnarray}
    \label{dyn_eq}
       \nabla_{\lambda}\left(
       \frac{1}{\mathcal{B}(x,\phi)}
       \,\sqrt{|K(\Gamma,\phi)|}\, (K^{-1})^{\mu\nu} \right)=0.
    \end{eqnarray}
where  for ease of notation we defined  the quantity
\begin{eqnarray}
\label{k}
K_{\mu \nu}(\Gamma,\phi) \equiv  \kappa^{-1}[\textbf{R}_{\mu\nu}(\Gamma)-\nabla_{\mu}\phi\nabla_{\nu}\phi] -\mathcal{A}(x,\phi)\textbf{u}_{\mu}\textbf{u}_{\nu}  \nonumber \\
   \end{eqnarray}
This simply means that the connection which has been taken arbitrary in the action is reduced now to the Levi-Civita connection of an invertible rank-two tensor $g_{\mu\nu}$ satisfying the condition
\begin{eqnarray}
\label{eom2}
   \sqrt{|g|}\,g^{\mu\nu}\equiv \frac{1}{\mathcal{B}(x,\phi)}\,\sqrt{|K(\Gamma,\phi)|}\, (K^{-1})^{\mu\nu}.
\end{eqnarray}
Indeed, here we should note  that, this tensor is meaningful as long as $K_{\mu\nu}$ is invertible and $\mathcal{B}(x,\phi) \neq 0$. The first requirement is not restricted to the present theory, it is in fact at the heart of most purely affine theories {\textit{\`a}} \textit{la} Eddington \cite{eddington}. On the other hand, the second requirement implies that in the matter-free case, a nonzero cosmological term has to replace $\mathcal{B}(x,\phi)$ leading to a theory with a cosmological constant. This shows that unlike metric gravity, the purely affine theories of gravity demand definitely a nonzero vacuum energy \cite{affine inflation, eddington}.

Thus, this tensor, which has not been imposed from the beginning, is now  generated dynamically through the variational principle and will play the role of the metric tensor. In fact, equation (\ref{dyn_eq}) is now equivalent to
\begin{eqnarray}
   \nabla_{\lambda}(\sqrt{|g|}g^{\mu\nu})=0.
\end{eqnarray}
It is important to notice however that concerning the signature of this metric, only those configurations $(\phi,\Gamma)$ where the tensor field (\ref{k}) has the signature $(-,+,+,+)$ are considered. This is a fundamental issue concerning the signature of the generated metric not only here in scalar-connection gravity but in all Eddington's type of purely affine theories \cite{kijowski}. In other words, one must consider only those configurations $\phi$ and $\Gamma$ where the tensor field $K_{\mu\nu}$ has negative determinant to guarantee that the generated metric tensor has a Lorentz signature. 
However, generally speaking, it seems that a more reasonable and convincing mechanism for fixing the metric signature has to be explored. In fact, unlike GR, the metric now is related to the dynamical connection and matter, which may suggest a varying signature too. Although the contexts are different, we mention here that in quantum field theories it has been shown that spacetime signature could gain also a dynamical character, and likewise, several attempts have been considered to devise a mechanism for a dynamical origin of the physical (Lorentzian) signature \cite{signature in qft}. That being said, we believe that finding a way that guarantees a Lorentz signature whether within the action itself (in terms of matter and curvature) or by providing a dynamical origin for it, would certainly reveal more interesting features of the affine gravitational theories.

Returning to the previous equations, the curvature $\textbf{R}_{\mu\nu}(\Gamma)$ becomes the Ricci tensor $R_{\mu\nu}(g)$ constructed by the metric $g$. Hence, equation (\ref{eom2}) represents the gravitational field equations, and with (\ref{k}) it takes the form
\begin{eqnarray}
\label{grav_eom_minimal case}
    R_{\mu\nu}(g) =&& 8\pi\left[\mathcal{A}(x,\phi)\textbf{u}_{\mu}\textbf{u}_{\nu}
    +\mathcal{B}(x,\phi)g_{\mu\nu} \right]
    + \nabla_{\mu}\phi\nabla_{\nu}\phi \nonumber \\
\end{eqnarray}
where we have used units $G_{N}=1$.

In terms of Einstein tensor, and by taking $\textbf{u}^{\mu}\textbf{u}_{\mu}=-1$, the above  equation reads
\begin{eqnarray}
\label{einstein equations}
R_{\mu\nu}-\frac{1}{2}g_{\mu\nu}R=&& 8\pi\left[\mathcal{A}\,\textbf{u}_{\mu}\textbf{u}_{\nu}+\left(\frac{1}{2}\mathcal{A}-\mathcal{B} \right)g_{\mu\nu} \right] \nonumber \\
&&+\nabla_{\mu}\phi\nabla_{\nu}\phi 
-\frac{1}{2}\,g_{\mu\nu} \nabla^{\lambda}\phi \nabla_{\lambda}\phi.
\end{eqnarray}

An important remark drawn here is that, besides the scalar field, these equations coincide with Einstein field equations sourced by a matter fluid with the stress-energy tensor
\begin{eqnarray}
T^{\text{m}}_{\mu\nu}= \mathcal{A}\,\textbf{u}_{\mu}\textbf{u}_{\nu}+\left(\frac{1}{2}\mathcal{A}-\mathcal{B} \right)g_{\mu\nu},
\end{eqnarray}
which implies that in the particular case where the fluid parameters are independent of the scalar field,  $\mathcal{A}(x,0)=\rho +P$ and $\mathcal{B}(x,0) =(\rho -P)/2$ with $\rho$ and $P$ being energy density and pressure respectively. Note that when $\phi$ is constant, the cosmological constant case corresponds to $\mathcal{A}=0$ which renders the SCG action (\ref{min_action}) to be the Eddington's action \cite{eddington}.

Next, we vary the action (\ref{min_action}) with respect to the field $\phi$ and  use the expression of the metric (\ref{eom2}) to find  
\begin{eqnarray}
\label{box phi}
\Box \phi=
4\pi \left(
\frac{\partial \mathcal{A}(x,\phi)}{\partial \phi}\, g^{\mu\nu}\textbf{u}_{\mu}\textbf{u}_{\nu} 
+2 \frac{\partial \mathcal{B}(x,\phi)}{\partial\phi}\right).
\end{eqnarray}
Indeed, since the matter parameters are $\phi$-dependent in the first place, the system will then develop potential-like terms. Needless to say, in the absence of matter this equations will simply describe a propagating massless scalar field. As we will see in the following section, equation (\ref{box phi}) is the basis of the phenomenon of spontaneous scalarization in the framework of SCG.

\section{Spontaneous scalarization}
\label{sec:spontaneous scalarization}
Most of the works that have been performed on spontaneous scalarization are based on the scalar-tensor theories of gravity and their modifications. The main essence behind this is usually the coupling of matter to a rescaled metric that incorporates the scalar field. However, as we have seen from the previous section, the metric tensor in the scalar-connection theory of gravity arises only \textit{a posteriori} through the dynamical equations, thus, the matter sector had to be coupled to the scalar field without metric in the action through its parameters $\mathcal{A}$ and $\mathcal{B}$.

Apart from the gravitational action (\ref{min_action}) which reflects the conceptional difference between SCG and STG, the obvious way to distinguish between the two theories is through the dynamics of the scalar field. In (\ref{eom in STG}), the right-hand side term acts as a potential within the trace of the energy-momentum tensor of matter which is given explicitly in terms of energy density and pressure \textit{a priori}. In (\ref{box phi}), however, the fluid parameters must be given \textit{a posteriori} by respecting the GR-limit. Nevertheless, in what follows we will show that SCG would lead to the same physics as STG in particular cases.

\subsection{Damour-Esposito-Far\`{e}se model from SCG}
Although the scalar-connection gravity can be considered as an alternative theory of gravity different from scalar-tensor theory, it might be relevant to produce the latter for certain cases. In fact, based on the previous remarks on the limit of the theory (GR), one may choose the quantities $\mathcal{A}(x,\phi)$ and $\mathcal{B}(x,\phi)$ to have following forms
\begin{eqnarray}
\label{a}
&&\mathcal{A}(x,\phi)=(\rho+P)e^{\beta \phi^{2}/2}, \\
&&\mathcal{B}(x,\phi)=\frac{1}{2}(\rho-P)e^{\beta \phi^{2}} 
\label{b}
\end{eqnarray}
where the scalar field-dependence of these parameters is chosen such that the GR solution arises at $\phi=0$. Spontaneous scalarization arises when this solution is unstable and the system then develops a stable solution at nonzero field. Here, to linear order the small perturbation $\delta \phi$ around GR solution leads to
\begin{eqnarray}
\label{box eq old}
\Box \delta \phi \simeq 4\pi \beta \Big[ 
(\textbf{u}^{\mu}\textbf{u}_{\mu}+2)\,\rho +(\textbf{u}^{\mu}\textbf{u}_{\mu}-2)\,P 
\Big]\delta \phi
\end{eqnarray}
For perfect fluids where $\textbf{u}^{\mu}\textbf{u}_{\mu}=-1$, the case that we have taken for the gravitational field equations (\ref{einstein equations}), we obtain
\begin{eqnarray}
\label{STG-equivalent}
\Box \delta \phi \simeq -4\pi \beta\left(-\rho +3P \right)\delta \phi,
\end{eqnarray}
which coincides with the Damour-Esposito-Far\`{e}se model (\ref{eom in STG}) where in this case the right-hand side of (\ref{STG-equivalent}) is given in terms of the trace of the energy momentum tensor $T_{\mu\nu}=(\rho+P)\textbf{u}_{\mu}\textbf{u}_{\nu}+P g_{\mu\nu}$. Hence, the physics of the scalar-connection theory is equivalent to that of scalar-tensor theory of gravity as long as the fluid parameters behave like (\ref{a}) and (\ref{b}). 

As one notices, the fluid parameters $\mathcal{A}(x,\phi)$ and $\mathcal{B}(x,\phi)$ respectively are not dependent on the scalar field similarly (here different exponents). This property can be traced back to the structure of the SCG Lagrangian where matter is described by two parameters instead of only one function (Lagrangian of matter) as in STG. To illustrate this fact, let us return to the general framework of the scalar-connection gravity formulated starting from action (\ref{min_action}). There, the fluid parameters described by the functions $\mathcal{A}(x,\phi)$ and $\mathcal{B}(x,\phi)$ were considered not only coordinate dependent but also $\phi$-dependent quantities. This implicit coupling between the scalar field and matter has led us to propose a relevant form for the fluid parameters as in (\ref{a})-(\ref{b}) allowing for the GR-limit for $\phi=0$. However, one may show that these parameters may enter the action, anyway, as $\phi$-independent functions $\tilde{\mathcal{A}}(x)$ and $\tilde{\mathcal{B}}(x)$ respectively. In this case, the SCG action may take the following form
\begin{widetext}
\begin{eqnarray}
S[\varphi,\Gamma]=\int d^{4}x \frac{\sqrt{|\kappa^{-1}\left[\bm{\omega}(\varphi)\textbf{R}_{\mu\nu}(\Gamma) -\nabla_{\mu}\varphi\nabla_{\nu}\varphi\right] -\tilde{\mathcal{A}}(x)\textbf{u}_{\mu}\textbf{u}_{\nu}|}}{\tilde{\mathcal{B}}(x)} 
\end{eqnarray}
\end{widetext}
where $\bm{\omega}(\varphi)$ is an arbitrary function of the scalar field $\varphi$ and represents the nonminimal coupling function in the scalar-connection theory of gravity.

An interesting feature of this action is the fact that the nonminimal coupling function can be absorbed easily without geometric transformation as in STG (by using metric transformation.) Here, this is realized by performing a simple field-redefinition of the scalar field as
\begin{eqnarray}
\nabla_{\mu}\phi\nabla_{\nu}\phi= \bm{\omega}^{-1}(\varphi)\nabla_{\mu}\varphi\nabla_{\nu}\varphi,
\end{eqnarray}
so that the last action reads
\begin{equation}
\label{transformed action}
S[\phi,\Gamma]=\int d^{4}x \frac{\sqrt{|\tilde{K}(\Gamma,\phi)|}}{\bm{\omega}^{-2}(\phi)\tilde{\mathcal{B}}(x)},
\end{equation}
in which
\begin{eqnarray}
\label{transformed kinetic term}
\tilde{K}_{\mu\nu}(\Gamma,\phi)=&&\kappa^{-1}\left[\textbf{R}_{\mu\nu}(\Gamma) -\nabla_{\mu}\phi\nabla_{\nu}\phi\right] \nonumber \\  &&-\bm{\omega}^{-1}(\phi)\tilde{\mathcal{A}}(x)\textbf{u}_{\mu}\textbf{u}_{\nu}. 
\end{eqnarray}
This minimal coupling picture of SCG described by action (\ref{transformed action}) tends to be analogues to (\ref{min_action}) where matter (fluid) is coupled to the scalar field. Now, we notice that the parameter $\tilde{\mathcal{A}}(x)$ describing the kinetic term of matter is rescaled by $\bm{\omega}^{-1}(\phi)$, whilst the potential-like term of the matter is recaled by $\bm{\omega}^{-2}(\phi)$. Hence, since the matter ($\phi$-independent) parameters $\tilde{\mathcal{A}}(x)$ and $\tilde{\mathcal{B}}(x)$ leads to the energy momentum tensor $T_{\mu\nu}=(\rho +P)\textbf{u}_{\mu}\textbf{u}_{\nu} +Pg_{\mu\nu}$, this explains now our choice of the parameters (\ref{a})-(\ref{b}) so that the scalar-connection theory leads to Damour-Esposito-Far\`{e}se model. Next, we will investigate the general case in which the present framework based on SCG produces a quite different results.   

\subsection{The general setup}
Here we will not follow the previous mechanism based on the transition from minimal to nonminimal coupling which leads to specific rescaling of the fluid parameters via (\ref{transformed action}) and (\ref{transformed kinetic term}). Rather, the general setup is based on the fact that the matter parameters $\mathcal{A}$ and $\mathcal{B}$ can generally depend on $\phi$ in different ways. One generic case is to take
\begin{eqnarray}
\label{a in terms of phi}
&&\mathcal{A}(x,\phi)=(\rho+P)e^{\beta_{1} \phi^{2}/2}, \\
&&\mathcal{B}(x,\phi)=\frac{1}{2}(\rho-P)e^{\beta_{2} \phi^{2}}
\label{b in terms of phi}
\end{eqnarray}
where unlike the previous case, now one has $\beta_{2}\neq \beta_{1}$, and the linearized equation of motion of the scalar field reads
\begin{eqnarray}
\label{final1}
\Box \delta \phi \simeq 4\pi\left[\left(2\beta_{2}-\beta_{1} \right)\,\rho - \left(2\beta_{2}+\beta_{1} \right)\,P  \right] \delta \phi, 
\end{eqnarray}
or
\begin{eqnarray}
\label{final2}
\Box \delta \phi \simeq -4\pi (\beta_{1}-2\beta_{2})\rho \left[1+\frac{\beta_{1}+2\beta_{2}}{\beta_{1}-2\beta_{2}}\omega(\rho) \right]\delta \phi
\end{eqnarray}
with $\omega(\rho)=P/\rho$ being the equation of state of matter. 

Thus, the spontaneous scalarization mechanism in this class of SCG is to be determined in line with specific bounds of the two parameters $(\beta_{1},\beta_{2})$. Given this different structure in terms of energy density and pressure compared to (\ref{STG-equivalent}), one may merely point out various scenarios for instance when $\beta_{1}=-2\beta_{2}$, the effective potential that drives spontaneous scalarization arises only through the energy density of matter. However, in this case the complete suppression of the instability will not be quite similar through both parameters $\mathcal{A}(x)$ and $\mathcal{B}(x)$ due to the different signs of the constants $\beta_{1}$ and $\beta_{2}$. The second scenario is when the dynamics is driven only by pressure, this arises here when $\beta_{1}=2\beta_{2}$ thus
\begin{eqnarray}
\Box \delta \phi \simeq -8\pi \beta_{1}P \delta \phi.
\end{eqnarray}
In this case, if $\beta_{1}$ is taken negative which is a trivial choice to eventually suppress the unstable modes then a negative effective mass square $-8\pi \beta_{1}P$ requires $P<0$. This cannot describe a physical high density object such as a neutron star unless the negative sign is conventional, i.e, the inward gravitational pull.

Nevertheless, a neutron star can be approximated to a nonrelativistic matter where the energy density term dominates in (\ref{final1}), hence 
\begin{eqnarray}
\label{linearized eq}
\Box \delta \phi \simeq 4\pi\,\left(2\beta_{2}-\beta_{1} \right)\rho \, \delta \phi. 
\end{eqnarray}
Again, for $\beta_{2}<\beta_{1}/2$ spontaneous growth occurs when the GR solution $\phi=0$ is unstable due to the negative mass squared $m^{2}_{\text{eff}}=4\pi \left(2\beta_{2}-\beta_{1} \right)\rho$ with an effective wavelength
\begin{eqnarray}
\lambda_{\text{eff}}=\sqrt{\frac{\pi}{|2\beta_{2}-\beta_{1}|\rho}}.
\end{eqnarray}
A sufficient approximation of the energy density of a star in terms of its mass $M$, radius $R$ and compactness $\mathcal{C}=M/R$ is $\rho \sim M/R^{3}$ which yields 
\begin{eqnarray}
\lambda_{\text{eff}} \simeq \frac{R}{\sqrt{|2\beta_{2}-\beta_{1}|\mathcal{C}}}.
\end{eqnarray}
Finally,  for  scalarization  to  occur,  this  effective  wavelength must fit inside the star, i.e, $\lambda_{\text{eff}}<R$.  The last condition implies a lower bound on the compactness of the star as
\begin{eqnarray}
\mathcal{C} \gtrsim \frac{1}{|2\beta_{2}-\beta_{1}|}.
\end{eqnarray}

Although it appears similar to familiar spontaneous scalarization with sole constant $\beta$ as in (\ref{STG-equivalent}), here the two nonzero constants reflect the fact that suppression of the eventual instability
in scalar-connection gravity must occur through two parts of matter; the kinetic part (\ref{a in terms of phi}) given in terms of the inertial gravitational mass density, and the potential-like part (\ref{b in terms of phi}).

In order to fit various observational constraints such as those coming from the pulsar-white dwarf binary \cite{observation}, it might be necessary even for SCG to embody a massive scalar field. In fact, one may easily show that in this case, instead of (\ref{linearized eq}) the system would gain an effective mass as  
\begin{eqnarray}
m^{2}_{\text{eff}} =  
4\pi \left(2\beta_{2}-\beta_{1} \right)\rho + m^{2}_{\phi},
\end{eqnarray}
where $m_{\phi}$ is now the mass of the scalar field $\phi$, and we have considered again the nonrelativistic scenario in which the energy density dominates in the effective potential.

The massive case tends to be important on cosmological scales where it prevents the entire universe at early times from scalarization \cite{in cosmology,fethi}. One possible way to constraint the scalar field mass has been used where the field wavelength associated to this field is taken much smaller than the so called periapsis orbit, i.e. $\lambda_{\phi} \ll r_{P}$ leading to $m_{\phi} \gg 10^{-16}$ eV \cite{fethi}. In this case however, we must have $\lambda_{\text{eff}} < \lambda_{\phi}$ which results in
\begin{eqnarray}
|2\beta_{2}-\beta_{1} | \mathcal{C} >m^{2}_{\phi}R^{2}.
\end{eqnarray}

We should assert here that these results are valid only for the specific classes that we have considered above, namely the particular form of the coupling of matter to the scalar field as well as the direct use of the the energy density of matter without referring to any equations of state of the compact objects. To that end, in order to accurately probe the SCG, one mainly has to examine various classes of the theory by providing a generic phenomenological study of other relevant forms different than (\ref{a in terms of phi}), (\ref{b in terms of phi}) and adopting specific equations of state $P=w(\rho)$ of neutron stars \cite{future work}. Last but not least, besides scalarization, it might be reasonable to accomplish other spontaneous growth mechanisms driven by distinct fields like fermions and vector fields \cite{fethi2} or through a Gauss-Bonnet term \cite{gauss bonnet}. However, this can be challenging since the structure of the gravitational actions that will be inherited from (\ref{min_action}) will not be obvious for those cases.

\section{Concluding remarks}
\label{sec:concluding remarks}
The work done in this paper aims to provide a scalar-connection gravity where unlike the famed scalar-tensor theories, gravity is not mediated by the metric tensor \textit{a priori}. Instead of that, the gravitational field is characterized by an affine connection as the fundamental element behind spacetime curvature. In the absence of metric, the simplest generally covariant action one can form is the celebrated Eiddington's action using only curvature \cite{eddington}. In this paper we have extended this simple action by introducing a scalar sector to mediate gravity through its coupling to spacetime connection. As in any theory of gravity, the gravitational actions must indeed embody ordinary matter to complete the gravitational dynamics. In this respect, putting various matter fields in actions formed only by the connections not the metric, has always been an exhausting challenge \cite{affine gravity with matter}. The reason is simply that field kinetic terms generally necessitate the physical metric in the first place to get their covariant form. Nevertheless, in forming the action of the scalar-connection gravity, we have proposed a handy method to invoke matter as a fluid in terms of one-form velocities. We have then shown that the resulting gravitational equations and the scalar field dynamics are obviously different from those of general relativity, the fact that indicates measurable deviations from the latter. 

In the second stage of the paper, we have performed a direct application of this framework by studying the mechanism of spontaneous scalarization as a relevant example. By making certain choices of the perfect fluid parameters that lead to GR solution for vanishing scalar field, we have seen that small perturbations around this unstable solution grows up due to an induced effective potential proportional to both energy density and pressure of matter. Moreover, we have shown that the theory leads to an equivalent prediction of scalar-tensor theory, namely the Damour-Esposito-Far\`{e}se model of spontaneous growth, in particular cases.  Furthermore, we have performed a theoretical study of the effects of the resulting effective potential on a compact star and discussed some bounds on our parameters in terms of the compactness of a star within which the latter is subject to scalarization.

Finally we should emphasize that although we have considered it as a relevant example in this paper, spontaneous scalarization must not be the only way to probe scalar-connection gravity. The latter is expected to lead to rich phenomenology when it is furthered in various aspects in astrophysics and cosmology.

\section*{Acknowledgments}
The authors are thankful to Fethi Ramazano\u{g}lu for discussing spontaneous scalarization. They are also grateful to Durmu\c{s} Demir and Gonzalo Olmo for discussing various aspects of purely affine gravity. This work is supported by the United Arab Emirates University (UAEU) research fund, UPAR Grant No. 31S434.

\end{document}